# The photonic wheel: demonstration of a state of light with purely transverse angular momentum

Peter Banzer[1,2,*], Martin Neugebauer[1,2], Andrea Aiello[1,2], Christoph Marquardt[1,2], Norbert Lindlein[2], Thomas Bauer[1,2], Gerd Leuchs[1,2]

[1] Max Planck Institute for the Science of Light, Guenther-Scharowsky-Str. 1, D-91058 Erlangen, Germany

[2] Institute of Optics, Information and Photonics, University Erlangen-Nuremberg, Staudtstr. 7/B2, D-91058 Erlangen, Germany

*Correspondence to: peter.banzer@mpl.mpg.de

The concept of angular momentum is ubiquitous to many areas of physics. In classical mechanics, a system may possess an angular momentum which can be either transverse (e.g., in a spinning wheel) or longitudinal (e.g., for a fluidic vortex) to the direction of motion. Photons, however, are well-known to exhibit intrinsic angular momentum which is longitudinal only: the spin angular momentum defining the beam polarization and the orbital angular momentum associated with a spiraling phase front. Here we show that it is possible to generate a novel state of light that contains purely transverse angular momentum, the analogue of a spinning mechanical wheel. We use an optical nano-probing technique to experimentally demonstrate its occurrence in our setup. Such a state of light can provide additional rotational degree of freedom in optical tweezers and optical manipulation.

In everyday life one often encounters spinning objects like the wheels of a running bicycle or a spiraling seed falling from a tree (*1*). Despite their huge differences in size, shape and material composition, a rolling wheel and a spiraling seed share an important property: both display a rotatory motion quantifiable via the angular momentum (AM) **J**. The latter is purely transverse for a rolling wheel that rotates around an axis perpendicular to the direction of motion of the bicycle, while it is entirely longitudinal for the spiraling seed which rotates around an axis parallel to the direction of fall.

The concept of AM lies at the foundation of physics of matter and waves. The AM of a massive system can be either transverse or longitudinal to the direction of motion of the system itself, as seen above. Remarkably, when it comes to angular momentum of light, a typical massless wave system, the situation is not as versatile. This is because sophisticated group theory arguments show that a photon with an exactly determined linear momentum **p** can only possess a longitudinal AM either parallel or antiparallel to **p** (*2, 3*). In fact, photons may exhibit both a longitudinal spin AM (the so-called helicity) (*4*), that defines the polarization of the light, and a longitudinal orbital AM (*5-7*), that sets the vorticity of the light. Thus, as far as the behavior with respect to AM is concerned, a photon is dynamically closer to a falling seed rather than a bicycle wheel. At the same time, there is not a no-go theorem that forbids the existence of more general states of light with purely transverse AM. Nevertheless, the existence of a photonic state analogue to a rolling wheel, as opposed to a spiraling seed, has eluded experimental verification up to now.

We use an optical nano-probing technique (*8, 9*) to demonstrate the occurrence of a state of light possessing a purely transverse AM, namely a "photonic wheel" (see Fig. 1), in the focal plane of a high numerical aperture (NA) focusing system. The appearance of this novel state of light is intimately connected with the recently discovered spin Hall effect of light (SHEL) (*10-15*) implied by transverse AM. In our setup a beam of light is first laterally split in two parts with opposite spin direction ($\sigma = -1$, left-circularly polarized and $\sigma = +1$, right-circularly polarized) and then tightly focused by a microscope objective (see Fig. 2 A and B). This yields to the occurrence of a purely transverse AM in the focal plane of the objective, as schematically illustrated in Fig. 2 B. We note here that similar experimental schemes were already proposed by (*16-18*), but for very different types of measurements and studies.

The electric field (EF) of a beam of light in the focus of a lens can be calculated by using the standard vectorial diffraction theory (*19, 20*). The results of these calculations for our laterally split input beam are depicted in Fig. 3 A - D. It may be seen that due to interference of the two halves of the beam with different spin, the y- and z-components of the EF show a pronounced shift in positive y-direction in the focal plane, while the x-component is shifted in the opposite direction. Notice that all EF components are of comparable strength. Moreover, the y- and z-components are $\pi/2$ out of phase which causes the spinning of the EF around the transverse x-axis, as shown in Fig. 1. On the optical axis of the focusing system ($x = y = 0$) the EF is linearly polarized in x-direction.

Finally, the barycenter of the total EF energy density distribution $|\mathbf{E}_{tot}|^2$ exhibits a shift in positive y-direction which is a manifestation of the geometric SHEL (*14, 15*) in the given scheme due to the transverse component $J_x$ of the beam AM. Concurrently, no spiraling phase front and thus no longitudinal OAM of one of the electric field components is observed in the focal plane (see insets in Fig. 3 A - C) which would be typical for a homogeneously circularly polarized beam that undergoes spin-to-orbital AM conversion by focusing (*21- 23*). This cancellation is pictorially illustrated in Fig. 2 B and calculated in Fig. 3 E - G where the focal distributions of the components of the AM density $\mathbf{j} = (j_x, j_y, j_z)$ are shown. By integrating these distributions over the focal plane, the AM per unit length $\mathbf{J} = \iint \mathbf{j}\,dxdy = (J_x, J_y, J_z)$ can be obtained. It follows that $J_y = J_z = 0$ while $J_x \neq 0$ because it exhibits an asymmetric distribution along the y-axis (see Fig. 3 H). This theoretically demonstrates that the AM is purely transverse in the focal plane, thus revealing the existence of a novel state of light, the "photonic wheel". It is important to note here that we can actually realize such a beam (with $J_z = 0$ while e.g. $J_x \neq 0$) not only in the focal plane. When leaving the focal plane, the longitudinal AM density $j_z$ is non-zero anymore but presents an anti-symmetric distribution that after integration results in a vanishing $J_z$.

In our experiments a special optical element is utilized to generate the tailored polarization and spin distribution which is then focused tightly (see Fig. 2 A and B). The element consists of two identical quarter-wave plates, merged with their fast-axes being oriented under +45° and -45° with respect to the bonding axis. With an appropriate choice of the polarization direction of the linearly polarized input beam, the desired beam is generated with spatially separated left- and right-handed circular polarization. Instead of using a linearly polarized $TEM_{00}$ mode as an input, we use a linearly polarized $TEM_{10}$ mode (see Fig. 2 A and B) to minimize diffraction at the split region between the quarter-wave plates. Please note that the basic concept of generating a state of light with purely transverse AM is not affected by this choice. The resulting optical beam is then tightly focused by a microscope objective with an NA of 0.9 (see Fig. 2 B) generating the aforementioned "photonic wheel". We use a scanning technique employing a gold nano-particle as a field-probe to experimentally proof the aforementioned statements about the focal shape of $|\mathbf{E}_{tot}|^2$ for the generated state of light with purely transverse AM. The single nano-particle is scanned through the beam in the focal plane (see Fig. 2 C). For every position of the nano-particle relative to the beam, the sub-wavelength particle (diameter 90 nm) interacts with the local electric field only. By detecting the light being scattered by the particle as a function of its positions relative to the focused beam in the focal plane, we are able to probe the focal distribution of $|\mathbf{E}_{tot}|^2$ (*20*). In the setup the forward-scattered and transmitted light is collected with an immersion-type microscope objective with an NA of 1.3 and guided onto a photodiode (see Fig. 2 C; see also (*20*)). For the measurements we choose a wavelength of 530 nm close to the plasmonic particle resonance. The employed experimental setup is similar to the one we already described in (*24*).

In Fig. 4 A we show our experimental scan result in transmission. A drop in signal strength is observed, when the particle is overlapping with the beam in the focal plane (see supplementary online text for further results). A smile-like deformation is observed experimentally for the specially tailored input polarization distribution described above. For comparison Fig. 4 B displays again the theoretically calculated distribution of $|\mathbf{E}_{tot}|^2$ for the experimental parameters. The experimentally measured shape of the beam in the focus is in excellent agreement with our theoretical predictions. Slight asymmetries in the experimental result are caused by deviations of the particle shape from an ideal sphere, and by imperfections of the split-aperture etc. From a theoretical point of view, the shape of the experimental result measured in the focal plane (Fig. 4 A) in combination with the chosen input polarization state is directly connected to our beam of light being in a purely transverse AM state in the investigated plane ($J_z = 0$).

As an application of this novel state of light with a purely transverse AM the aforementioned system can be implemented in an optical tweezer setup. A trapped particle should rotate around a transverse axis in this scheme in contrast to conventional AM transfer resulting in a rotation around the longitudinal axis. When combined with spin and orbital AM in the focal plane (*25-28*), such an implementation paves the way to three-dimensional rotation control over a trapped particle in an optical tweezer setup.

**Acknowledgments:** We thank M. Schmelzeisen from the Max Planck Institute for Polymer Research in Mainz, Germany for fabricating the spherical nanoparticles used in our experiments.

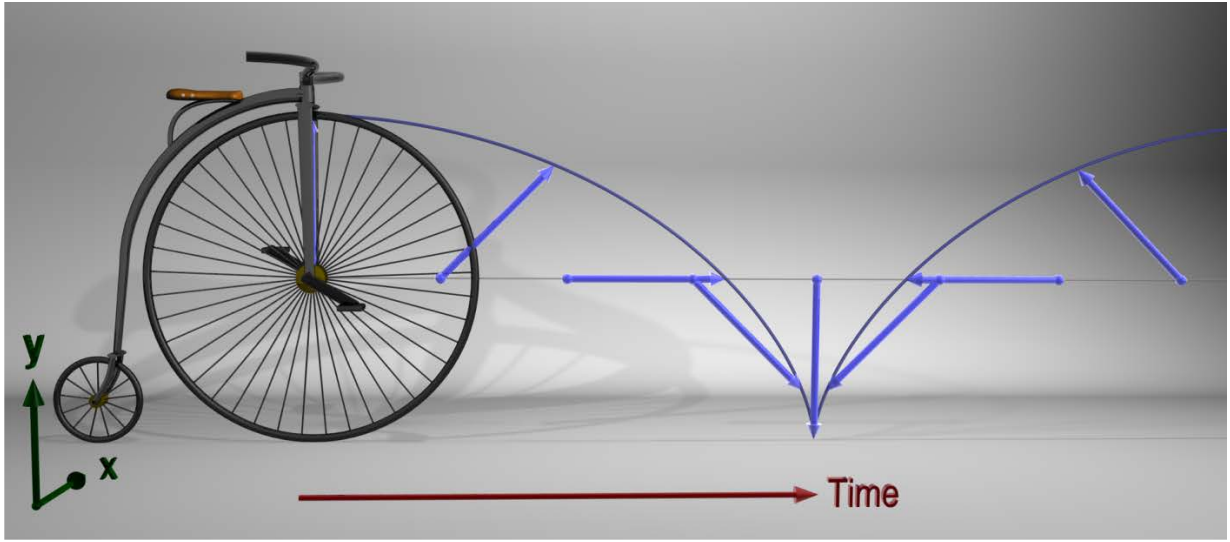

**Fig. 1**. Artistic illustration of a photonic wheel in analogy to a rolling mechanical wheel. The horizontal axis represents the time. The x- and y-axis span the focal plane for z = 0. The end tip of a single spoke of the bicycle's wheel traces a cycloidal curve in time. The same holds true for the electric field vector in a photonic wheel, which spins around the transverse coordinate axis x in the focal plane with time. The shown arrows represent the yz-components of the electric field vector at defined positions in the focal plane for a state of light with purely transverse AM.

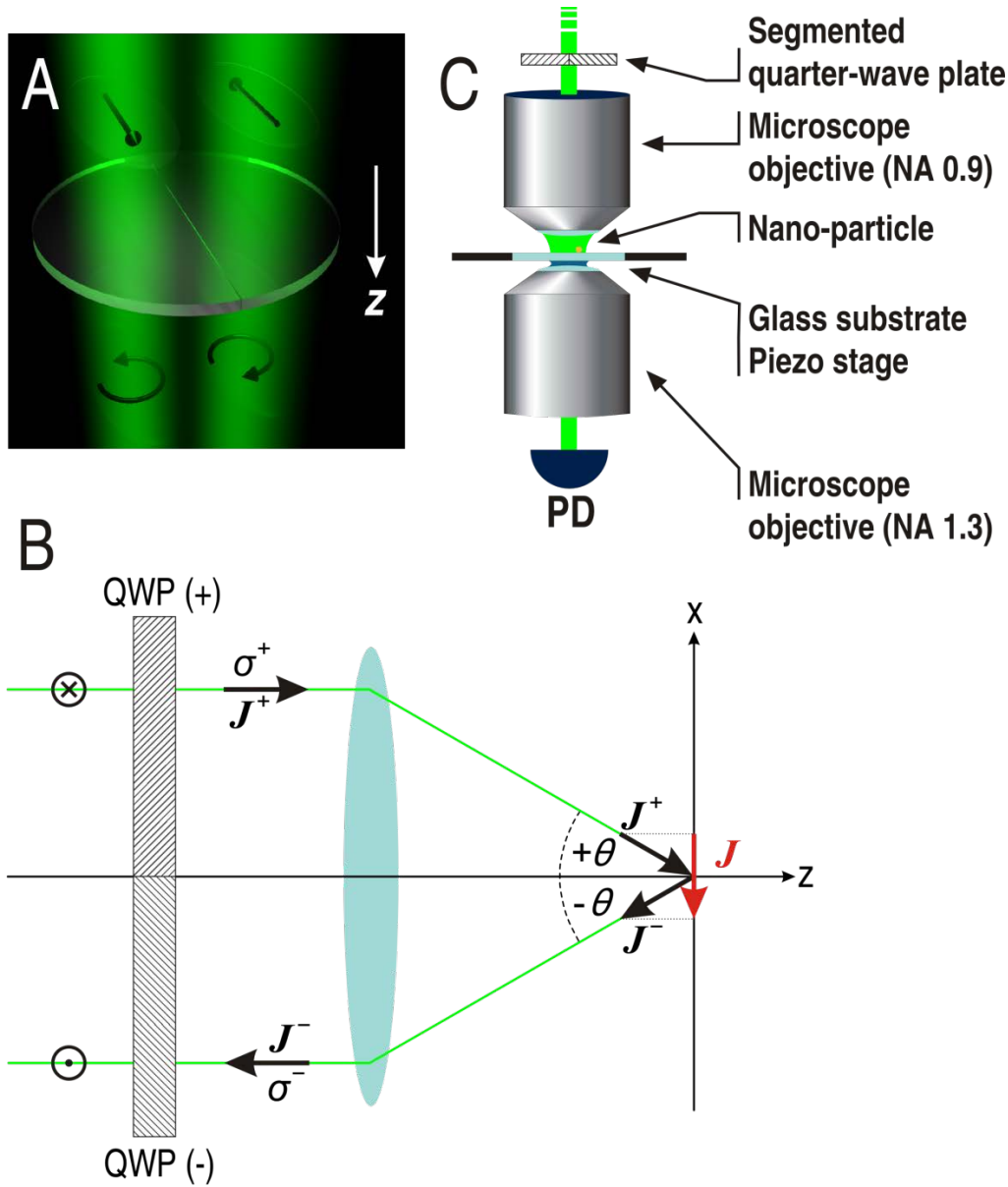

**Fig. 2.** (A) Optical element for generating the spin tailored input beam consisting of two laterally bonded quarter-wave plates (QWP) with orthogonal fast axes. A linearly polarized input beam (here a y-polarized TEM$_{10}$ mode) is transformed into a split beam with separated left- and right-handed circular polarization. (B) When tightly focused, the longitudinal components of $J^+$ and $J^-$ cancel, while their transverse components add up to a purely transverse AM (red) in the focal plane. (C) Experimental setup and measurement scheme. The state of purely transverse AM is generated in the focus of a microscope objective with an NA of 0.9. A gold nano-particle (diameter 90 nm) sitting on a glass substrate is scanned through the beam in the focal plane using a piezo-stage. For every position of the particle in the beam the forward scattered and transmitted light is collected by an immersion-type microscope objective (NA = 1.3) and the corresponding intensity is measured with a single photodiode.

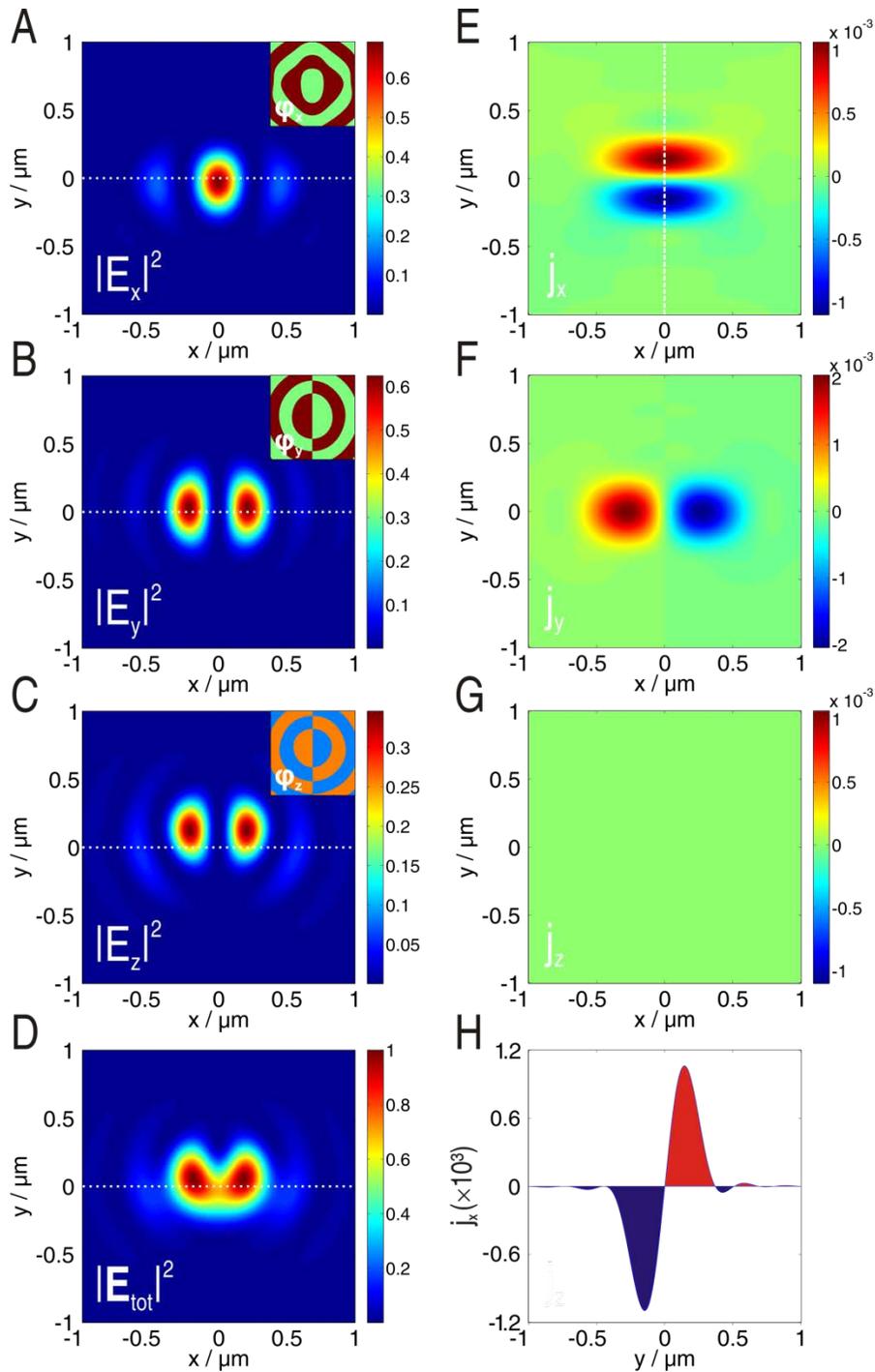

**Fig. 3.** (A) - (D) Theoretical distributions of $|E_x|^2$, $|E_y|^2$, $|E_z|^2$ and $|\mathbf{E}_{tot}|^2$ for a helicity tailored input beam in the focal plane (normalized to max ($|\mathbf{E}_{tot}|^2$)). Insets: corresponding distributions of the relative phases $\phi_x$, $\phi_y$ and $\phi_z$ (dark red: $\pi$; light green: 0; orange: $+\pi/2$; light blue: $-\pi/2$). (E) - (G) Calculated distributions of the components $j_x$, $j_y$ and $j_z$ of the AM density $\mathbf{j}$ confirming the existence of purely transverse AM in the focal plane. (H) Cut through the asymmetric distribution of $j_x$ along the y-axis as indicated by the dashed white line in (E).

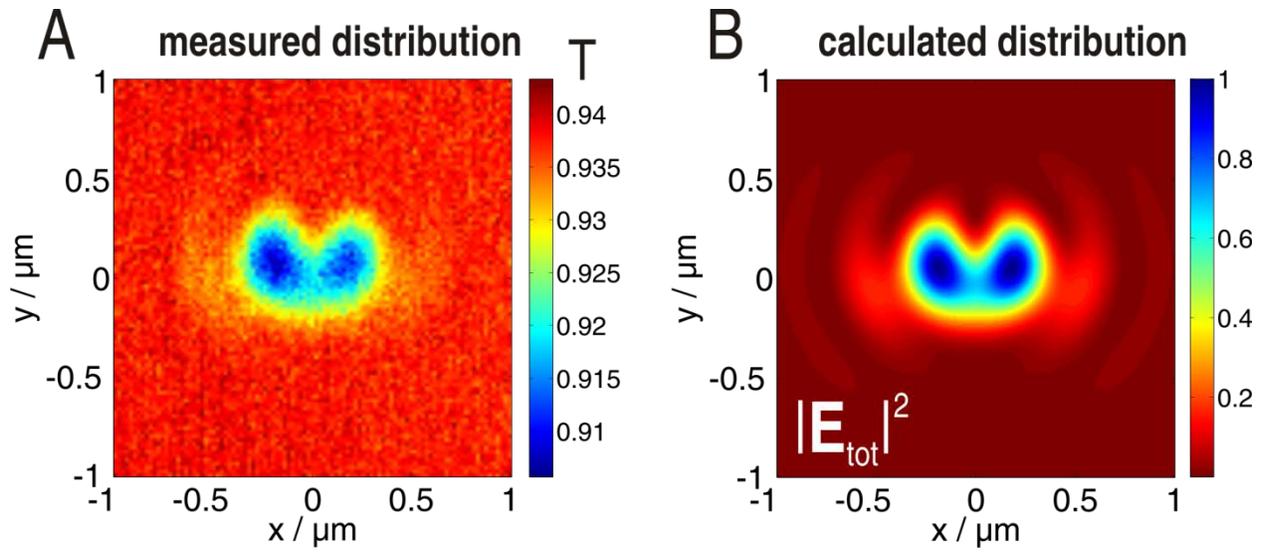

**Fig. 4.** (A) Experimental scan results (wavelength: 530 nm; normalized transmission T is shown). Every pixel of the presented 2D scan image corresponds to one position of the nano-particle relative to the beam in the focal plane. The integrated transmitted and forward scattered intensity for each position is shown. The experimental data is normalized to the transmission of the focused beam through the glass substrate. (B) Corresponding calculated distribution of $|\mathbf{E}_{tot}|^2$ (shown with inverted color-map for better comparability).

## Supplementary Materials:

## Materials and Methods:

### Simulation

To calculate the focal field distributions using the Debye integrals (*19*), the incoming beam is sampled on a grid of points in the entrance pupil of the focusing system. At each sampling point the beam is locally treated as a geometrical ray being refracted by the lens system taking into account its state of polarization and propagating towards the geometrical focus with an individual **k**-vector. In the focal plane, these rays represent local plane wave components which interfere with each other forming the focal field distribution. The integrals are solved numerically. For calculating the presented theoretical electric energy density and angular momentum density distributions in the focal plane of an aplanatic system, a total number of 200 × 200 plane waves was used for the decomposition of the input beam. The presented area of the focal plane of 2 μm × 2 μm was sampled with a total number of 400 × 400 data points.

### Sample fabrication

The spherical shape of the single gold nano-particles is achieved by treating conventional and commercially available gold nano-particles with intense laser pulses (see for instance (*29*)). This procedure was performed at the Max Planck Institute for Polymer Research (MPIP) in Mainz, Germany. The particles are molten by the high localized energy of the light pulses. Due to surface tension, the particles reform to an almost perfectly spherical shape. The nano-particles are dispersed in water and have a diameter of approximately 90 nm. Marker structures are patterned on the glass substrate (BK7) before the particles are spin-coated. To this end, laser-lithography is used to locally expose a photoresist on the substrate. After development, a 90 nm thick gold film is deposited. Finally, the unexposed photo-resist and the part of the metal layer on top of the resist are removed by a lift-off procedure. The marker structures are utilized for orientation and precise focus alignment purposes in the experiment. Furthermore they allow for an exact characterization of the employed nano-particle.

### Experimental setup

In the experimental setup the light beam from a super-continuum light-source (NKT Photonics/Koheras SuperK PowerPlus) is spectrally filtered using an acousto-optical tunable spectral filter (type: AA Opto-Electronic MDSnC-TN AOTF) operating in the visible spectral range. Thereupon, the beam is coupled into a single-mode fiber for mode-cleaning purposes. Behind the fiber the linearly polarized Gaussian-shaped beam is collimated again. Before being tightly focused, the polarization and intensity distribution of the collimated beam is transformed into the required distribution using the segmented quarter-wave plate (custom-built by B. Halle Nachfl. GmbH) as already described in the main manuscript. If the input beam is converted to a Hermite-Gaussian mode of odd order (for instance a linearly polarized $HG_{10}$ mode) before passing through the segmented wave plate, diffraction originating from the split region between both wave plates can be drastically reduced. Furthermore, the maxima of the intensity distribution of the input field are shifted towards the outer rim of the entrance aperture. Therefore, on the one hand stronger z-components are generated in the focal plane relative to the transverse field components. On the other hand, also stronger transverse AM is created in the focal plane leading to a more pronounced shift of the barycenter of the electric energy density distribution. The theoretical as well as experimental results presented in this article were achieved using such a linearly polarized $HG_{10}$ mode as an input beam for the segmented quarter-wave plate. The prepared beam is then coupled into a microscope objective with a high numerical aperture of 0.9 (type: Leica HCX PL FL 100×/0.90 POL 0/D). The sample which we use for beam-profiling consists of single and well separated spherical nano-particles made from gold with a diameter of approximately 90 nm sitting on a glass substrate, which can be positioned relative to the focal plane with nanometer resolution using a 3D piezo-stage. The light scattered off the single nano-particle is collected by an immersion-type microscope objective in transmission with an NA of 1.3 (type: Leica HCX PL FLUOTAR 100×/1.30 OIL). The substrate is index-matched to the microscope objective's front lens using immersion oil. The transmitted light is detected by means of a single photodiode. The employed experimental setup is similar to that we already described in (*24*).

**Reconstruction of the focal electric energy density distribution using a single gold nano-particle on a substrate**

Let us consider the basic case of a point-like electric dipole oscillating harmonically in free space or inside a homogeneous medium. It is well known that the emitted power has a $\sin^2(\theta)$ distribution with θ being the angle relative to the dipole moment. By defining the z-axis as the optical axis of an optical system collecting the emitted power, two special cases can be investigated: the dipole moment perpendicular (transverse) or parallel (longitudinal) to the z-axis. If in both cases the dipole moment is of the same strength, a full hemisphere along the positive or negative z-axis has to be collected to measure the same integrated power in both scenarios, which is half of the total power emitted. If instead a limited solid angle (NA) is used for collection, the integrated collected power will be mismatched. Equivalently, if for instance a single spherical nano-particle embedded in free-space or a homogeneous medium is used to interact with the electric energy density distribution in the focal plane of a high NA optical system, the same considerations have to be taken into account. If the particle is chosen to be sufficiently small, its dipolar response dominates. Here, the nano-particle is excited by the local electric field, which is almost homogeneous across the particle diameter. The direction of the induced electric dipole moment then just depends on the direction of the locally linear electric field it is overlapping with. When collecting the forward scattered and transmitted and the backward-scattered light in this configuration, a mismatch between the collected integrated power is observed for a longitudinally or transversely oscillating electric dipole which was induced in the particle with equal strength. Furthermore, in forward direction also the transmitted part of the incoming beam which has not interacted with the particle and which can interfere with the forward-scattered part of the light field is present. If in a next step either the electric point-like dipoles or the scattering particle are located close to a dielectric interface the situation changes significantly. For instance, as a consequence of the glass substrate used as a support for the nano-particle in the actual experiment, its resonance is spectrally red-shifted relative to the free-space case, which is taken into account in the experiments by choosing a proper wavelength of excitation. Beside the red-shift of the spectral resonance position induced by the presence of a dielectric interface, also the emission pattern of a point-like electric dipole or the excited nano-particle is changed drastically. Although the particle is small in diameter and the excitation is locally linear, the emission is not purely dipole-like anymore and the emitted pattern in the glass hemisphere differs from that in air (*30-32*). Hence, the integrated emissions into the air and glass hemisphere are not equal in this case. It can be shown, that in such a situation collecting the forward scattered light within a solid angle of less than $2\pi$ is sufficient to measure the same power emitted by electric dipoles excited in the particle by equally strong longitudinally or transversely oscillating electric fields. Thus, the energy densities of the corresponding field components can be reconstructed correctly by using a nano-particle as a field probe. Furthermore, the transmitted light which has not interacted with the particle and which can interfere with the forward scattered light changes the angular dependence slightly. In the experimental system we used for the measurements the transmitted and forward scattered light is collected with a microscope objective with an NA of 1.3. The corresponding collection angle within the glass hemisphere is larger than the angle discussed above for which longitudinal and transverse field components are probed equally. Nevertheless, only a slight mismatch between the reconstructed longitudinal (z) and transverse (x and y) components of the electric energy density is expected. Thus, for confirming the deformed shape of our beam in the focal plane, the chosen and simplified scheme of measurement is sufficient. It is worth mentioning here that we will report on a highly accurate reconstruction of amplitude and even phase of the individual electric field components of tightly focused and arbitrarily polarized beams by taking into account the aforementioned details elsewhere soon.

**Supplementary Text:**

**Further experimental results**

In addition to the presented experimental results, we confirm that the experimentally observed shape of the beam in the focal plane is not caused by any aberration of our optical system but is indeed a direct consequence of the spin distribution in the entrance aperture. For that purpose, we first used an x-polarized Gaussian input beam which is tightly focused. It is known that due to the vectorial character of the light field the focal distribution of $|\mathbf{E}_{tot}|^2$ should be elongated along the axis of polarization (*19, 33*). For comparison, the corresponding theoretical distribution is presented in Fig. S1 B. The experimental scan image is in good agreement with the theoretically predicted shape and the elongation along the x-axis is clearly visible in Fig. S1 A. In Fig. S1 C the experimental result for the spin tailored input beam is presented again ("photonic wheel"). Figure S1 D shows the corresponding focal electric energy density distribution.

**Generation of "photonic wheels" with different focal electric field distributions**

For trapping sub-wavelength nano-particles and an application of the presented scheme in optical tweezers, the freedom of changing the focal electric energy density distribution while preserving the purely transverse character of the AM is of great importance. Therefore, we also investigated a second input field distribution which generates a state of purely transverse AM in the focus, but exhibits a fundamentally different distribution of $|\mathbf{E}_{tot}|^2$ and its components. For that purpose, the relative phase between the left- and right-handed circular polarizations in the two beam segments of the input beam is changed (additional phase shift of $\pi/2$). The investigated input distribution is presented in Fig. S2 (left). The corresponding experimental result and the calculated total electric energy density distribution are shown in Fig. S2 A and B. Clearly, the focal distribution has changed quite drastically. This is because, for instance, the maximum electric energy density of the z-component of the electric field is now located on the y-axis also exhibiting a pronounced shift along this axis relative to the origin (not shown here). Furthermore, the maximum of the y-component of the electric energy density is found to lie on the y-axis as well. The z- and y-component of the electric field are $\pi/2$ out of phase, which is in line with the predicted purely transverse angular momentum. In addition, the x-component of the electric field has a zero crossing on the y-axis. In this configuration the distributions of the AM density components are equivalent to those already presented in the main manuscript, thus still yielding $J_y = J_z = 0$ and $J_x \neq 0$.

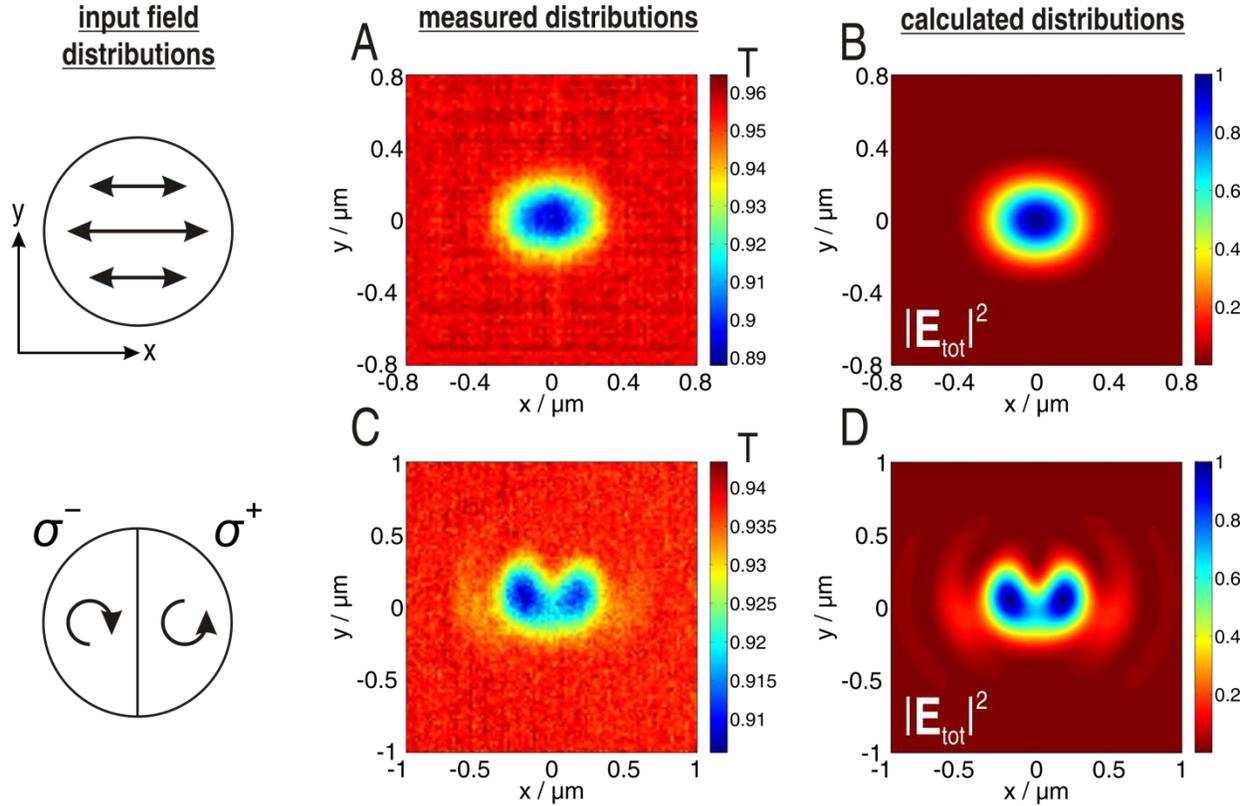

**Fig. S1.** (A) Experimental scan result (wavelength: 530 nm; normalized transmission T is shown) for a tightly focused and in x-direction linearly polarized input beam. Every pixel of the presented 2D scan image corresponds to one position of the nano-particle relative to the beam in the focal plane. The integrated transmitted and forward scattered intensity for each position is shown. The experimental data is normalized to the transmission of the focused beam through the glass substrate. (B) Corresponding calculated distribution of $|\mathbf{E}_{tot}|^2$ (shown with inverted colormap for better comparability). (C) Experimental result for a "photonic wheel" (compare to main manuscript) and corresponding theoretical distribution of the total electric energy density.

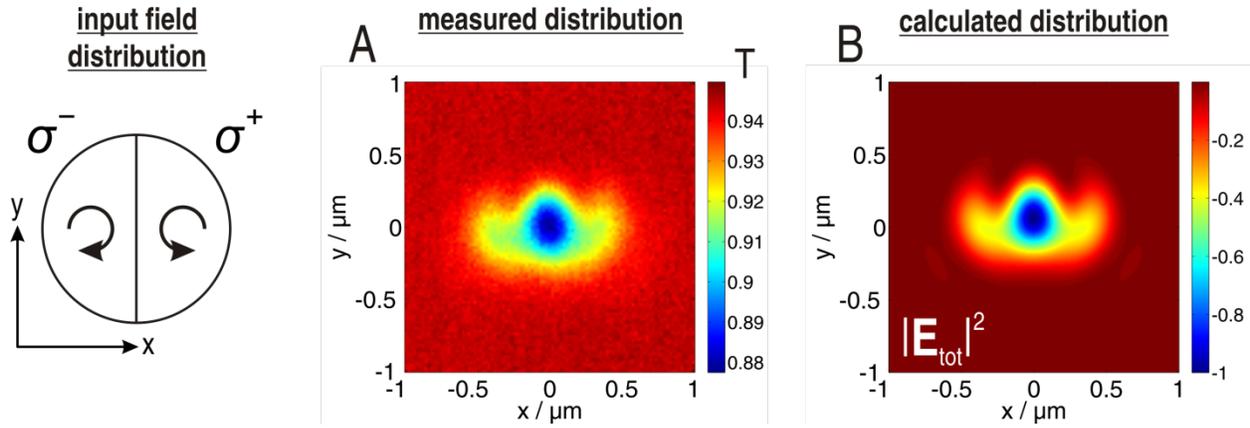

**Fig. S2.** A) Experimental scan result (wavelength: 530 nm; normalized transmission T) for an alternative realization of a "photonic wheel". The two halves of the input beam exhibit an additional phase shift of $\pi/2$ in comparison to the case before. (B) Corresponding calculated distribution of $|\mathbf{E}_{tot}|^2$ (shown with inverted color-map for better comparability).